# Transient Phenomena in Sub-Band Gap Impact Ionization in Si NIPIN Diode

Bhaskar Das[1], Jörg Schulze[2] and Udayan Ganguly[1]

*Abstract*—Sub-band-gap (SBG) impact ionization (II) enables steep subthreshold slope that enable devices to overcome the thermal limit of 60mV/decade. This phenomenon at low voltage enables various applications in logic, memory and neuromorphic engineering. Recently, we have demonstrated sub-0.2V II in NIPIN diode experimentally primarily based on steady state analysis. In this paper, we present the detailed experimental transient behavior of SBG-II in NIPIN. The SBG-II generated holes are stored in the p-well. First, we extract the leakage mechanism from the p-well to show two mechanisms (i) recombination generation (RG) and (ii) over the barrier (OTB) where OTB dominates when barrier height $\phi_b < 0.59 eV$. Second, we analytically extract the SBG II current ($I_{II}$) at 300K from experimental results. The drain current ($I_D$), electric field ($E-field$), and $I_{II}$ are plotted in time. We observe that $I_{II}$ increase as $E-field$ reduces which indicates that *E-field* does not primarily contribute to $I_{II}$. Further, the $I_D$ shows two distinct behaviors (i) $I_{II}(I_D)$ is constant at the beginning and (ii) eventually *"universal"* $I_{II}(I_D)$ is linear, i.e. $I_{II} = k * I_D$ where k=$10^{-3}$; We also show that the electrons primarily contributing to $I_D$ are directly incapable of II due to insufficient energy ($< E_g$). Fischetti's model showed that SBG-II is primarily caused by "hot" electrons that accept energy in an Auger-like process from "cold" drain electrons to enable SBG-II. We speculate that if the $I_D$ electrons "heat-up" the cold drain electrons, which would further energize the hot electrons to produce the observed $I_{II}(I_D)$ universal dependence.

## I. INTRODUCTION

Sub-band gap impact ionization in devices generates steeper subthreshold slope than 60mV/dec. II has been used in selector diodes to have high on/off current ratio for RRAM applications [1], in memory programming [2- 5], in fast switching devices [6, 7], in neural devices for neuromorphic computing [8]. Several research groups [9-13] have experimentally demonstrated sub-band gap II in Si by measuring the body current in steady state (DC) condition in SOI MOSFETs that can be used for low voltage applications. Fischetti et al. [14] have shown by full band Monte Carlo simulation that the main cause of sub-band gap II in Si is the collision of source electrons to the *'cold'* drain electrons. They have shown that for sub 100$nm$ channel length devices with low channel doping, the transport is almost ballistic and collision happens mostly in the drain. If the energy of the source electron is less than the band gap energy ($< E_G$), that collision should not cause any impact ionization. However, they showed that, there is some finite probability for some higher energy drain electrons to receive this energy ($< E_G$) and make II possible. They have also calculated the II rate. Recently we have experimentally demonstrated Si based NIPIN diode [15] to detect lowest reported (till date) voltage for sub-band gap II in Si at $0.2V - 0.5V$ at 300K. In this paper, the transient behavior of II and the effect of electric field ($E-field$) and over the barrier current ($I_D$) is investigated experimentally for the first time. The DC and transient results are discussed in section II and III. The impact ionization current is calculated using a capacitor divider model [15] (section IV). The effect of $E-field$ and $I_D$, on sub-band gap bias II is discussed in section V.

## II. EXPERIMENT AND SIMULATION

The device structure, experimental doping profile (SIMS data), fabrication steps and the basic working principle of NIPIN diode is mentioned in the previous work [15]. A schematic of NIPIN diode is shown in fig.1(a). Here the same NIPIN diode, as in [15], is used only with top contact area is $200\ \mu m \times 200\ \mu m$ for all the measurements. TCAD simulation for NIPIN was performed using Sentaurus™ software, where only ideal drift diffusion (i-DD) condition with no impact ionization has been considered, as the DD simulation in Sentaurus over estimates the II [14 -16]. The details of simulation parameter and conditions are mentioned in [15]. DC (steady state) IV and transient (current vs time) measurements were performed to demonstrate the $E-field$ and $I_D$ effect on II in Si in sub-band gap biases.

### A. DC Measurements

The presence of II in NIPIN diode was established by comparing the experimental and simulated (w/o II) DC IV characteristics (fig. 1b) as mentioned in detail in [15].

### B. Transient Measurements

The transient measurements were performed on Agilent B 1500A waveform generator/fast measurement unit (WGFMU).

The authors[1], Bhaskar Das (email: bhaskardas@ee.iitb.ac.in), Udayan Ganguly (email: udayan@ee.iitb.ac.in) are with the Dept. of Electrical Engineering, Indian Institute of Technology Bombay.
The author[2] Jörg Schulze (email: schulze@iht.uni-stuttgart.de) are with University of Stuttgart, Germany.

Funding is acknowledged from Science and Engineering Research Board (SERB), Department of Science & Technology (DST), Dept. of Electronics and IT (DeitY), Govt. of India for partial support.

As significant amount of II is clearly observed for $V_a < 0V$ case (fig. 1b) [15], only negative applied biases ($V_a < 0V$) have been considered for further measurements. Square voltage pulses having rise time = fall time = $10ns$ (minimum limit of the measurement system) and a pulse width of $6000ns$, were applied with peak voltages varied from $-0.5V$ to $-0.7V$. Inset of fig.1 (c) shows one of the applied square input voltage pulse ($V_a = -0.65V$). The pulse width was chosen ($6\mu s$) such that for all the applied biases the transient current can reach its steady state value within this time. The output current versus time (I-t) for applied biases $-0.5V$ to $-0.7V$ is plotted in fig. 1(c). Due to the sudden rise (rise time $10ns$) of input voltage pulse, a spike is generated in the output (fig. 1c). This is because of displacement current due to capacitor charging [15]. To get the initial ($t < 200ns$) current without any II effect which is superposed with the spike, the transient output current has been extrapolated from the region ($\sim 10^{-6}s$) where there is no influence of the spike, to the point when the input square pulse just reaches its peak value ($t = 135ns$ – inset of fig. 1c), as shown in fig. 1(c). This current values at $t = 135ns$ (red circles in fig. 1 c) gives the without II current at each biases. The steady state current values for each applied biases are also marked (green dots) in fig. 1c. In fig. 1(d) these w/o II and steady state current values from transient measurements are compared with the experimental and simulated (w/o II) DC IV [15].

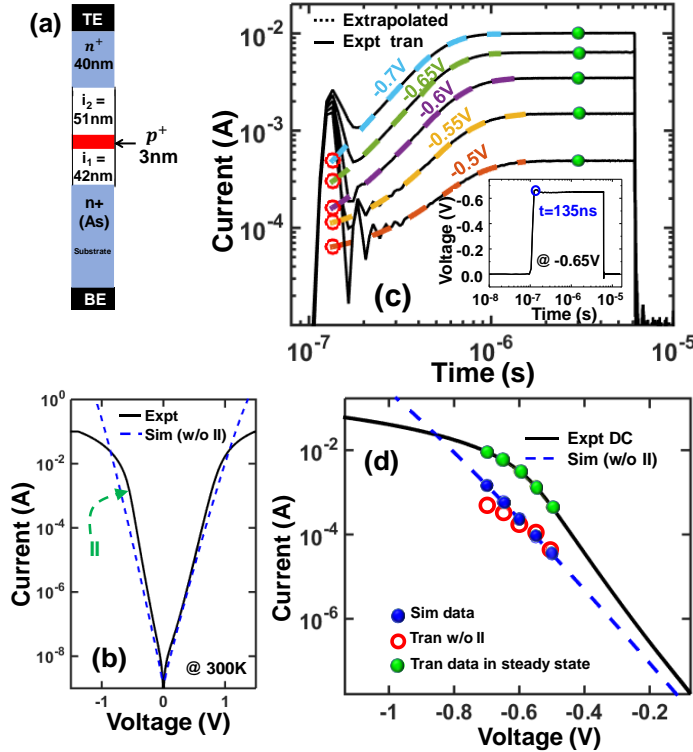

Fig.1(a) Schematic of NIPIN diode, (b) Experimental DC IV (continious black line) and simulated without II IV (blue dotted line), (c) Experimental transient current vs time (black continious line), extrapolated I-t (coloured dotted lines), inset – applied square voltage pulse (V-t) of peak voltage -0.65V. (d) Comparison of transient without II data (red circle) with simulated w/o II DC IV data (blue spheres) and transient saturated state current (green spheres) with experimental DC IV [15].

### III. RESULTS AND DISCUSSION

Fig. 1(c) shows that the transient current gradually increases from a lower value to its saturated value. Because, as the applied square pulses reaches its peak value (within rise time = $10ns$), the barrier of the device is lowered instantaneously and over the barrier current ($I_D$) flows from source to drain. With the increase in time, these carriers start to collide with the drain side electrons [14, 15] and some of them impact ionize. As the channel doping is very low ($10^{16}/cm^3$), there is almost no collision in the channel. Impact ionization generated holes are being stored in the well, which causes the barrier to become further low. Due to this barrier lowering, more OTB electrons makes II in the drain and more holes are being stored, which acts like a positive feedback mechanism. These stored holes are being lost either by 1) OTB hole diffusion or by 2) recombination. Eventually the equilibrium is reached when the hole generation due II equals the hole loss, and the current reaches its saturation value. From fig. 1(c), it is seen that, as the applied bias increases, the time taken by the current to reach its saturation level decreases. This might be because, with the increase of applied bias, the number of carriers (electrons) that overcomes the barrier increases exponentially that increases the probability of impact ionization, therefore, it takes less time for the well to fill with holes until the saturation state is reached and the current reaches the saturation state faster for higher biases.

These saturation current values (fig. 1c) at each applied biases should match with the DC current value (fig. 1b) at that applied bias. At $t = 3000ns$, the current ($I_D$) reaches its saturation value for all the applied biases. So, the current values at $t = 3000ns$ are taken and compared with the DC IV (green dots fig. 1d), and they matches exactly. At $t = 135ns$ the applied square pulse just reaches its peak value (inset of fig. 1c). Neglecting any delay between the input voltage pulse and the output current, the output current at $t = 135ns$ should give OTB current without any II, considering there will be no barrier lowering due to II generated hole storage in the well by this time. This extrapolated current values at $t = 135ns$ is compared with the simulated w/o II (no II parameter included in the simulation) data in fig. 1(d). At lower biases until $V_a = -0.55V$ these values matches pretty well with the simulated results, however as the bias increases, the mismatch increases. This is because of the effect of series resistance in the fabricated device. This series resistance comes into play for $V_a > -0.55V$ and the experimental DC current starts to saturate (fig. 1b). However, the simulated current is not affected by the series resistance in this bias regime as the series resistance of the simulated device is lower than the experimentally fabricated devices. The well agreement of simulation (w/o II) data with transient results (for $V_a < -0.55V$) validates that extrapolated drain current ($I_D$) at $t = 135ns$, gives over the barrier current without any presence of II. In the next section, we calculate the II current at different $V_a$ comparing these experimental w/o II current values and a capacitor divider model [15].

### IV. II CURRENT CALCULATION

The schematic band diagram of NIPIN at $V_a$ is shown in fig. 2(a). The red dashed line demonstrate the band with barrier height of $\phi_B$ just after the application of $V_a$ and the corresponding $I_D$ is w/o II current. The blue continuous line represent the band condition at any latter time with barrier height of $\phi_A$. The II generated holes is stored in the well and these holes leak away either by diffusion or by recombination. The current equation in the transient state (before the steady state is reached) can be written as

$$I_{II} = I_{h-diff} + I_{h-stored} \quad \text{(in transient state)} \quad (1)$$

where $I_{II}$ is the II generated current, $I_{h-diff}$ is the over the hole barrier diffusion current (fig. 2a) and $I_{h-stored}$ is the hole store current in the well. Here we have assumed that the recombination current is negligible. This assumption is proved with experiment in the latter part of this section. The steady state is reached when the hole generation due to II is equal to hole loss and then $I_{h-stored} = 0$ and equation (1) changes to

$$I_{II} = I_{h-diff} \quad \text{(in steady state)} \quad (2)$$

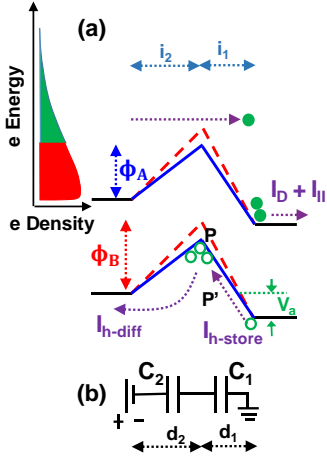

Fig.2(a). Schematic of NIPIN band diagram at $V_a$, (b) capacitor divider model.

Therefore, to calculate the $I_{II}$, we have to find both $I_{h-diff}$ and $I_{h-stored}$ and which is determined from the amount of stored holes in the well at each time ($t$). The stored holes in the well at any time ($t$) is calculated by capacitor divider model (fig. 2 b) [15]. The capacitor values are calculated using equation (3) and (4)

$$C_1 = \frac{\varepsilon \times Area}{d_1} \quad (3)$$
$$C_2 = \frac{\varepsilon \times Area}{d_2} \quad (4)$$

The hole stored at any time ($t$) in the well (position denoted by P in fig. 2a) is $h_P$.

$$h_P = \frac{(C_1+C_2) \times (\phi_B - \phi_A)}{q \times Volume} \quad (5)$$

The hole diffusion current at $t$ is given by equation (6). Where the hole concentartion at the position $P'$ (at the same energy level as the source side valance band) is given by equation (7). The hole stored current at t is calculated using equation (8). Here, $q$ is electronic charge, $D_P$ is diffusion constant for holes.

$$I_{diff} = Area \times (q \times D_P \times \frac{1}{d_2}$$
$$\times \frac{(C_1+C_2) \times (\phi_{w/o\ ii} - \phi_{with\ ii})}{q \times Volume}$$
$$\times e^{\frac{-q \times \phi_{with\ ii}}{kT}})$$
$$= Area \times (q \times D_P \times \frac{P'_{d_2}(h\ density\ at\ P')}{d_2}) \quad (6)$$

$$h_{P'} = h_P \times e^{\frac{-q \times \phi_A}{kT}} \quad (7)$$

$$I_{stored} = (\frac{h\ Stored\ at\ (t_1+1) - h\ Stored\ at\ (t_1)}{(t_1+1) - (t_1)} \times q \times area) \quad (8)$$

In the following part we prove the assumption of considering the recombination loss current to be negligible compare to diffusion hole loss.

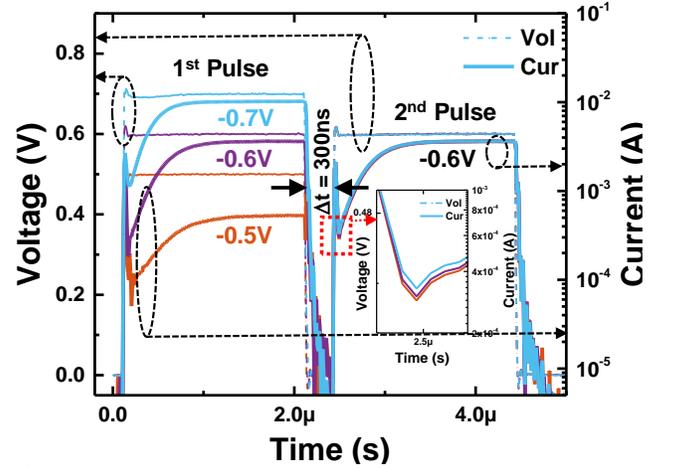

Fig. 3. Applied two square pulses (dotted line) and corresponding output currents (continuous line) vs time. For the first pulse, peak voltages are $V_a = -0.5V, -0.6V\ and -0.7V$ and for second pulse, the peak voltage is $V_a = -0.6V$ for all the cases. Inset shows the zoomed part of red square portion.

To prove this hypothesis, a transient measurement with two consecutive square pulses (rise time = fall time = $10ns$, pulse width = $2000ns$) of which the 1st one had peak voltages $V_a = -0.5V, -0.6V, -0.7V$ and the 2nd one had a fixed peak voltage of $V_a = -0.6V$ was performed. The time gap between these two pulses ($\Delta t$) was varied from $20ns$ to $1500ns$ (in steps) for all the three cases of applied bias. Fig. 3 shows the applied pulses for these three cases of 1st pulse of $V_a = -0.5V, -0.6V, -0.7V$ with a fixed 2nd pulse of $V_a = -0.6V$ having $\Delta t = 300ns$. Corresponding output currents are also plotted. The current ($I_D$), when the applied bias just reaches the peak voltage ($t_{2nd-0}$) for the 2nd pulse, was extracted by extrapolating (not shown in fig. 3) the current to the time $t_{2nd-0}$ in the similar method as mentioned earlier (fig. 1c). These extracted current values for all time gaps ($\Delta t$) is plotted with $\Delta t$ in fig. 4 (a). Fig. 4 (a) shows that there are two slopes for all the applied bias cases ($V_a = -0.5V, -0.6V, -0.7V$). These slopes are fitted with exponential fit. The possible mechanism for the observation of these two slopes is explained by using band diagrams (fig. 5). For a peak voltage of $V_a$ for the 1st pulse, the band diagram of NIPIN in steady state is shown fig. 5 (a). The barrier height is $\phi_e$ which is lower than the barrier height of $\phi_{w/o\ II}$ (barrier height in the absence of II), due to the stored holes in the well because of II. The band diagram at a time which is after the end of the 1st pulse and before the starting of the 2nd pulse is shown in fig 5 (b). As in the time gap $\Delta t$ no II is presence, some of the stored holes (due to the application of the 1st pulse), will escape from the well. As diffusion loss is barrier dependent and recombination loss is not, we speculate that initial loss process will be dominated by the diffusion process as the barrier is low due to stored holes. Due to this hole loss, barrier will go up with increase in $\Delta t$. After a 'critical time' ($t_c$), the barrier will reach a 'critical value' ($\phi_{hc}$) when the recombination process will start do dominate over the diffusion process because of the higher barrier. In fig. 5(b), $\phi_e < \phi_{hc}$, and $0 < \Delta t < t_c$, so in this time zone diffusion is dominant. On the other hand, when $\Delta t > t_c$ (fig. 5c), $\phi_e > \phi_{hc}$, therefore, recombination dominates. Finally at $t \rightarrow \infty$, (fig. 5d) all the stored holes are removed and the barrier becomes $\phi_e = \phi_{w/o\ II}$. Fig. 4(a) demonstrate this process. Over the barrier current ($I_D$) depends on (i) the barrier height which depends on the amount of hole stored and (ii) on $V_a$. For the second pulse (fig. 3) $V_a = -0.6V$ for all three cases of 1st pulses. The initial currents at the starting of the 2nd pulse (fig. 4a) is different because of different amount of stored holes for different $V_a$ applied in the 1st pulse (fig. 3). Comparing these initial $I_D$ values in fig. 4 (a) we find, it is obvious that the amount of hole stored due to 1st pulse for $V_a = -0.7V$ is higher than that for $V_a = -0.6V$ and $V_a = -0.5V$ cases. With the increase of $\Delta t$ the hole should leak away from the well and the barrier should increase and the initial current due to 2nd pulse $I_D$ should match for all three cases when the hole remained in the well is same for three cases and barrier height matches. Therefore, all the decay curves should align with each other if they are time shifted properly. To align the currents (fig. 4a), a time-shift was applied by $60 ns$ for $V_a = -0.5V$ and $20 ns$ for $V_a = -0.6V$ cases, as shown in fig. 4(b). As a result, all the curves fall on top of each other. From fig. 4 (b), it is clear that there are two slopes of these decay process (denoted by dotted lines). The decay time ($\tau$) for the initial slope (fig. 4b) is $\sim 20 ns$ and for the latter part it is $\sim 120 ns$. The average life time for hole recombination process for a doping of $2 \times 10^{18} - 3 \times 10^{18}/cm^3$ is approximately in the range of 200ns-400ns in Si [17-18]. So we speculate that the initial fast hole decay is due to diffusion process and the slow decay at 2nd part of the curve (fig. 4b) is mainly due to recombination process. From fig. 4(b), the calculated values are $t_c \sim 270 ns$ and $\phi_{hc} \sim 0.59 eV$. For all the II based hole stored measurements (fig. 1c), $\phi_e \ll \phi_{hc}$, so the assumption of neglecting recombination over diffusion is hence justified and equation (1) and (2) are valid. Fig. 6 shows the variation of three currents $I_{h-diff}$, $I_{h-stored}$ and $I_{II}$ with time calculated using equation (1), (6) and (8) for $V_a = -0.6V$. In the beginning (at $t = 135 ns$), there is almost no hole stored in the well, therefore, $I_{II}$ starts from a lower value and then reaches to a steady state value. After the II process starts, the holes begins to accumulate in the well, and the diffusion current ($I_{h-diff}$) starts to flow, which is clear from the fig.6. Initially $I_{h-stored}$ has a higher value and then gradually decreases with the decrease in barrier height. $I_{h-diff}$ increases with time and catches up $I_{II}$ in the steady state.

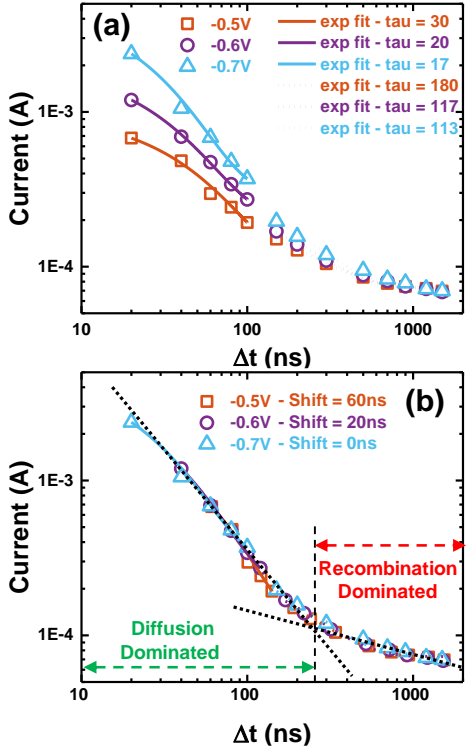

Fig.4. Output current at the beginning of the second pulse vs $\Delta t$ for $V_a = -0.5V, -0.6V$ and $-0.7V$ (for first pulse), (a) as extracted. Exponentially fitted lines are for diffusion-dominated region (continuous) and for recombination-dominated region (dotted), (b) shifted in time axis by $60 ns$ for $V_a = -0.5V$ and by $20 ns$ for $V_a = -0.6V$. Two slopes in the current (black dotted lines) is due to diffusion-dominated leakage and recombination-dominated leakage process.

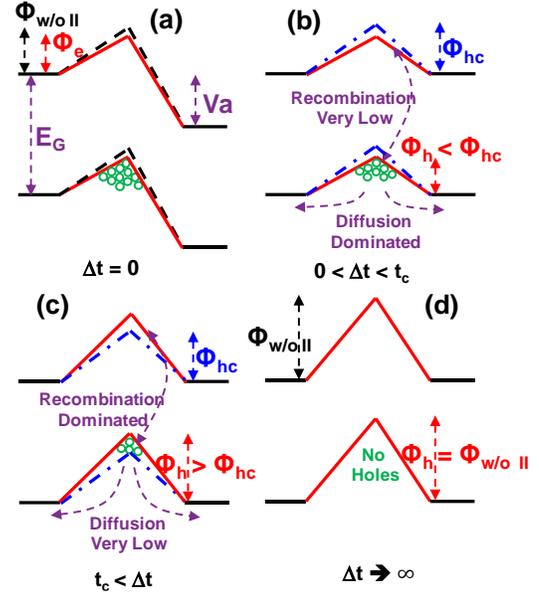

Fig. 5. Band diagram at the beginning of the second pulse (a) for $\Delta t = 0$, with II (continuous line) and without II (dotted line), (b) for $0 < \Delta t < t_C$, $\phi_{hc} = critical\ barrier$ (dotted line), $\phi_h = barreir\ due\ to\ stored\ holes$ (continuous line), (c) for $t_C < \Delta t$, $\phi_{hc} = critical\ barrier$ (dotted line), $\phi_h = barreir\ due\ to\ stored\ holes$ (continuous line), (d) for $\Delta t \rightarrow \infty$, $\phi_{hc} = \phi_h$.

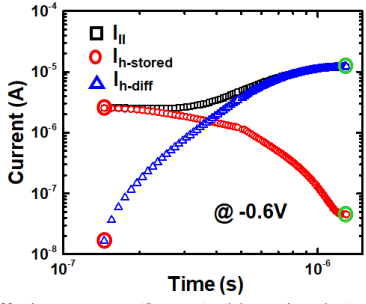

Fig.6. Hole diffusion current ($I_{h-diff}$) (blue triangles), current due to hole storage ($I_{h-stored}$) (red circles), impact ionization current ($I_{II}$) variation with time for applied bias -0.6V in log scale. Start point is marked by red circle and end point is matked by green circle.

## V. Effect of E-field and $I_D$ on Impact Ionization

The variation of $I_D$, barrier height ($\phi$) and $E-field$ with time for all applied biases are plotted separately in fig.7 (a-c). Corresponding $I_{II}$ vs time is calculated using the above mentioned method and plotted in fig. 7(d). From fig. 7(b) it is

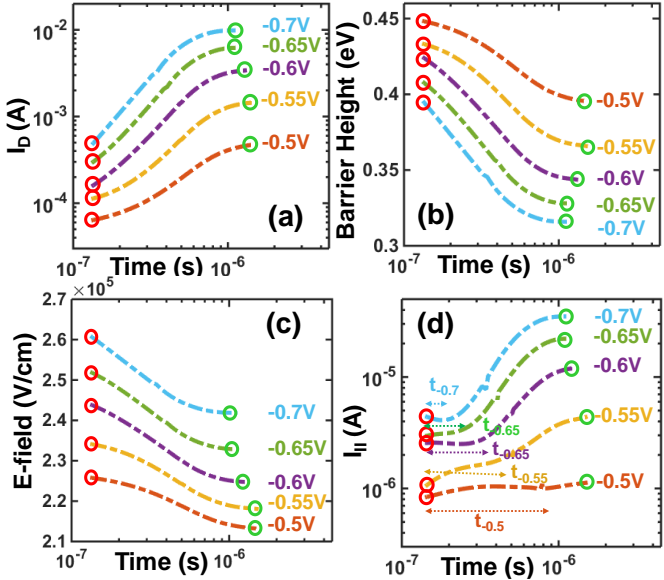

Fig.7 (a) drain current ($I_D$) vs time, (b) Barrier height vs time, (c)E-field vs time, (d) impact ionization current ($I_{II}$) vs time, the initial period of almost constant $I_{II}$ for each $V_a$ is marked by dotted arrow and denoted with $t_{V_a}$ for corresponding applied biases ($V_a$). Start point of every curve is marked by red circle and end point is matked by green circle.

clear that with the increase of time, barrier height decreases as the holes starts to accumulate in the well, and finally reaches the steady state value. Due to the decrease in barrier height, $E-field$ decreases (fig. 7c) and $I_D$ increase with time (fig. 7a) for all the $V_a$ ($-0.5V\ to\ -0.7V$). From fig. 7(d) it is observed that for $V_a = -0.5V$, $I_{II}$ remains almost constant with time, whereas, for $V_a > -0.55V$, $I_{II}$ initially remains almost constant (this time is indicated by $t_{V_a}$ at an applied $V_a$) and then increase with time. The variation of $I_{II}$ with $E-field$ (fig. 8a) and with $I_D$ (fig. 8 b) in the transient as well as in the steady state (from experiment) is shown in fig. 8.

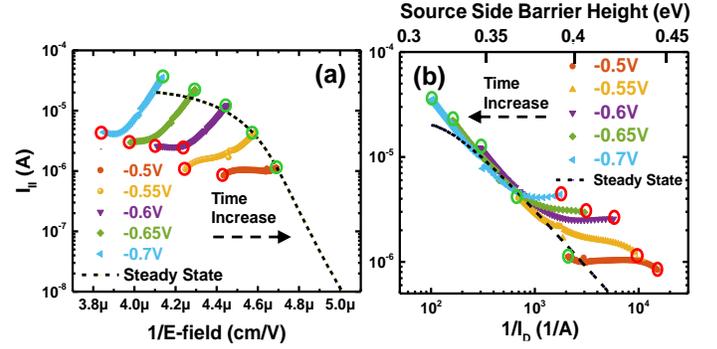

Fig.8 (a) In steady state (black dotted line), in transient state (colored lines) for applied biases $-0.5V$ to $-0.7V$, the variation of (a) impact ionization current ($I_{II}$) vs $E-field$ and the variation of (b) impact ionization current ($I_{II}$) vs 1/over the barrier drain current ($I_D$) and Source side barrier height ($\phi_{BS}$).

The current due to II ($I_{II}$) depends on two parameters (i) $E-field$ and (ii) number of carriers that overcomes the barrier ($I_D$). If $E-field$ increases, the energy gained by the carriers, before collision increases, which increases the probability of II. If the number of carriers increases, the probability of the number of carriers potentially capable of II also increases, so it increases II. Therefore, $I_{II}$ is some function of $E-field$ and $I_D$, which increases with $E-field$ and $I_D$. Here, in steady state (fig. 8 a,b), with the increase in $V_a$, as both $E-field$ and $I_D$ increases, $I_{II}$ increases. However, in the transient cases, the $E-field$ increases and $I_D$ decreases with time simultaneously. Therefore, the increase in $I_{II}$ with time is positively correlated to the increase in $I_D$ (fig. 8b) and negatively correlated to increase in $E-field$ (Fig. 8a). It appears that the $E-field$ dependence of $I_{II}$, which is valid for a *pn* junction with long depletion region with multiple $e-h$ pair generation per injected electron, does not hold for this case. We observe from fig.8 (a, b) that, $I_{II}$ correlates with $I_D$ (also related to source side barrier height $\phi_{BS}$) but not with $E-field$.

To understand this observation, we use the band diagram (fig 9a), where the electrons current from the source are exponentially dependent on source side barrier where the barrier cuts-off electrons below the $\phi_{BS}$ and only electron above $\phi_{BS}$ contribute to current. As the barrier height reduces with time, the current $I_D$ increases exponentially. We observe that the $I_{II} \propto k * I_D$ where, $k > 10^{-3}$ i.e. 1/1000 electron impact ionizes (fig. 7a, d). In addition, at these low energies, each electron at the most produce one $e-h$ pair during impact ionization. However, based on the band diagram, the source barrier ($\phi_{BS}$) reduction in time adds lower energy electrons (the pink band of electrons as shown in fig. 9a). These newly added electrons carry drain-barrier ($\phi_{BD}$) worth of energy which is lower than $E_G$ as shown in fig. 9(a-b). Hence, these electrons cannot directly impact ionize even though they can contribute to current. Essentially, the electron current capable of II is not modulated by the barrier modulation in time. Yet, we observe the $I_D$- dependence of $I_{II}$. To understand this, we review that Fischetti et al. [14] has proposed a mechanism of sub-bandgap impact ionization based on detailed Monte Carlo simulations. Here, the process of energy exchange between "*cold*" electrons in the drain occurs where the energy of injected "*hot*" electrons mostly reduces. However sometimes (rarely) the energy of "*hot*" electrons also increases in an Auger like process [14]. This enhances impact ionization. We speculate that the observed enhancement of $I_{II}$ related to $I_D$ increase as observed, may occur if the increased $I_D$ (though unable to II directly but) heats up the "$cold$" electrons which will transfer the energy to hot electrons to enhance $I_{II}$ to show a $I_D$ dependence.

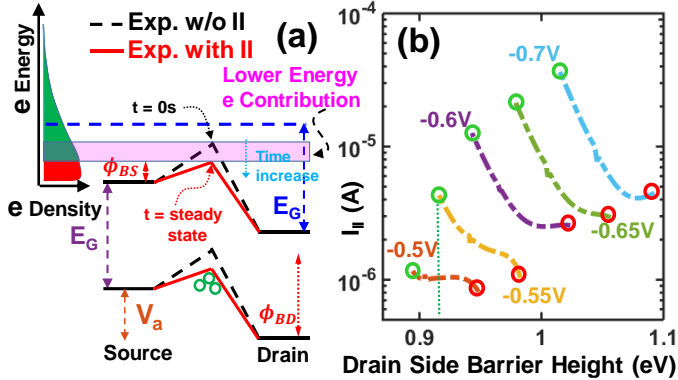

Fig. 9. (a) The schematic band diagram at $V_a$. $\phi_{BS}$ is the source side and $\phi_{BD}$ is the drain side barrier height at $V_a$. Electrons distribution in the source side having energy greater than $\phi_{BS}$ is marked by green and lower is marked by red. The barrier just after the application of $V_a$ is shown in black dotted line and the steady state barrier ($\phi_{BS}$) is shown by red line. Pink band of electrons are the added electrons to $I_D$ due to barrier lowering, (b) Variation of $I_{II}$ with drain side barrier height.

Initially, the transient $I_{II}$ shows an approximately constant $I_{II}$ (independent of $I_D$) which later joins the a "$universal$" $I_{II}$ vs $I_D$ dependence (fig. 8b). Intuitively, the observation that the constant, $I_D$-independent $I_{II}$ increases with $V_a$ is the consistent with increase in the initial drain barrier i.e. the energy of the electrons due to applied bias (Fig. 9(b)). It is observed that the specific time ($t_{Va}$) reduces at higher $V_a$. Essentially the time $t_{Va}$ is the time duration of the constant $I_{II}(I_D)$ regime when $I_D$ increases without change in $I_{II}$. The magnitude of the $t_{Va}$ is the time for the constant $I_{II}(I_D)$ regime to merge with the universal $I_{II}$ vs. $I_D$ dependence where $I_{II}$ is strongly related to $I_D$. While such a behavior has been observed for the first time, a physical explanation would require detailed Monte Carlo study from the experts in the community.

## VI. CONCLUSION

In this paper, the *NIPIN* device is studied to analyze the transient impact ionization process at sub-band-gap bias. The $h$-escape from the well is experimentally determined to show two mechanism (i) recombination-generation (RG) and (ii) over-the-barrier (OTB). We demonstrate a clear transition from RG to OTB current at $\phi_{BS} < 0.59 eV$. Based on the OTB hole loss model and $\phi_{BS}$ extracted from $I_D$, a simple method of extracting $I_{II}$ is developed. The transient $I_{II}$ has fixed bias but varying $I_D$ and $E-field$. We show that the $I_{II}$ has a counter-intuitive reduction with $E-field$, which indicates that $E-field$ is not the dominant contributor. On the other hand, $I_{II}$ dependence on $I_D$ has two strong regimes (i) constant $I_{II}(I_D)$ dependence which is $V_D$ dependent and (ii) a universal $I_{II}(I_D)$ dependence with a fixed $k = 1000$ i.e. every $1000^{th}$ electron contributing to $I_D$ produces an electron-hole pair by impact ionization. Our analysis indicates that with electron current producing $I_D$ modulated in time $I_{II}$ cannot directly produce $I_{II}$ as it has insufficient energy ($<E_g$). We invoke Fischetti's Auger like energy transfer of energy from "$cold$" drain electrons to "$hot$" injected electrons as a primary mechanism for sub-bandgap II to speculate that $I_D$ (once it achieves a sufficient extent) may heat up "$cold$" drain electrons which may enhance $I_{II}$. Thus, our experimental results present new observations on the electron dynamics studies during sub-bandgap bias impact ionization. On one hand, these studies should inspire physical modeling by Monte Carlo or other appropriate techniques. On the other hand, it is strong step towards low voltage II based devices for advanced computing applications